\documentclass{article}

\usepackage{arxiv}

\usepackage[utf8]{inputenc} 
\usepackage[T1]{fontenc}    
\usepackage{hyperref}       
\usepackage{url}            
\usepackage{booktabs}       
\usepackage{amsfonts}       
\usepackage{nicefrac}       
\usepackage{microtype}      
\usepackage{amsmath}
\usepackage{cleveref}       
\usepackage{upgreek}
\usepackage{bm}
\usepackage{diagbox}
\usepackage{graphicx}
\usepackage[numbers]{natbib}
\usepackage{doi}
\usepackage[version=4]{mhchem}

\title{Re-examining the boundary conditions in modelling SAW-driven acoustofluidic streaming}

\date{}

\author{ {Qinran Wei, Suyu Ding, Yang Zhao, Yuanpeng Ma, Dachuan Sang, Dong Zhang\thanks{dzhang@nju.edu.cn}, and Xiasheng Guo\thanks{guoxs@nju.edu.cn}} \\
	Key Laboratory of Modern Acoustics (MOE), School of Physics, \\
	Collaborative Innovation Center of Advanced Microstructures, Nanjing University,\\
	Nanjing 210093, China\\
}

\hypersetup{
pdftitle={Re-examining the boundary conditions in modelling SAW-driven acoustofluidic streaming},
pdfsubject={Acoustofluidics, Acoustic modelling, Nonlinear dynamics in fluids, Ultrasonics},
pdfauthor={Qinran Wei, Suyu Ding, Yang Zhao, Yuanpeng Ma, Dachuan Sang, Dong Zhang and Xiasheng Guo},
pdfkeywords={Acoustic streaming, Stokes drift, Surface acoustic wave, Boundary conditions},
}

\begin{document}
\maketitle

\begin{abstract}
Numerical simulations of surface acoustic wave~(SAW)-induced acoustic streaming are highly sensitive to the choice of second-order boundary conditions. This study systematically compares the no-slip~(NS) and Stokes slip~(SD) boundary conditions through different numerical approaches. Two- and three-dimensional simulations based on the Reynolds stress method are performed for standing SAW and travelling SAW devices. Results are validated against particle image velocimetry measurements of streaming patterns and velocities. We show that the SD condition yields Lagrangian velocity fields in significantly better agreement with experiments than the NS condition, accurately capturing vortex number, rotation direction, and amplitude across varying device geometries and operating conditions. In contrast, the NS condition overpredicts velocities by 1–2 orders of magnitude and often fails to reproduce experimentally observed vortex structures. These findings highlight the essential role of the Stokes drift boundary condition in modelling acoustic streaming and provide clear guidance for its use in future simulations of SAW-based acoustofluidic systems.
\end{abstract}


\section{Introduction}
As an emerging technique for manipulating micro- and nanoparticles, acoustofluidics offers excellent biocompatibility and holds significant potential for biomedical applications~\cite{zhang_acoustic_2020}. In acoustofluidic devices, particle behaviour is primarily governed by two second-order effects arising from acoustic waves---the acoustic radiation force~\cite{bruus_acoustofluidics_2012a, karlsen_forces_2015}, which is well-characterised, controllable, and relatively simple in form, and acoustic streaming~\cite{sadhal_acoustofluidics_2012, wiklund_acoustofluidics_2012}, which exhibits greater complexity and is more challenging to control. In many scenarios, these two effects exert competing influences on particles~\cite{muller_numerical_2012, muller_ultrasoundinduced_2013,qiu_particlesizedependent_2020,ni_modelling_2019,hoyos_controlling_2013,karlsen_acoustic_2018,bach_suppression_2020}.

Acoustic streaming has been the subject of extensive research. Lord Rayleigh pioneered the theoretical foundation and provided explanations for accompanying experimental observations~\cite{rayleigh1884circulation}. There are two fundamental types of acoustic streaming---boundary-driven Rayleigh streaming~\cite{rayleigh1884circulation} and Eckart streaming due to attenuation of the acoustic wave in the fluid~\cite{eckart_vortices_1948}. Seminal contributions to the understanding of both types have been made by Nyborg~\cite{nyborg_acoustic_1958,nyborg_acoustic_1965} and Lighthill~\cite{lighthill_acoustic_1978}.

The calculation of acoustic streaming dates back to Rayleigh, who provided an analytical solution for boundary-driven streaming generated by standing waves between infinite parallel plates~\cite{rayleigh1884circulation}. Hamilton~\textit{et al.}~\cite{hamilton_acoustic_2003} and Muller~\textit{et al.}~\cite{ muller_ultrasoundinduced_2013} further derived analytical expressions for boundary-driven streaming in channels with regular shapes and well-defined resonance modes. However, for acoustofluidic devices featuring complex geometries and vibration patterns, analytical solutions are often intractable, making numerical methods necessary.

Acoustic streaming in such contexts is commonly computed using acoustic perturbation theory~\cite{bruus_acoustofluidics_2012}, in which source terms derived from first-order acoustic fields are substituted into the time-averaged second-order equations to obtain the steady streaming flow. This methodology is also referred to as the Reynolds stress method (RSM)~\cite{lei_comparing_2017}. It is important to note that resolving the acoustic boundary layer is necessary, which typically requires highly refined meshes near all fluid-solid interfaces, leading to considerable computational demands~\cite{muller_numerical_2012}.

Regarding the boundary conditions for second-order acoustic streaming, several studies have employed a no-slip condition at the wall~\cite{nyborg_acoustic_1958,lee_nearboundary_1989,muller_numerical_2012,devendran_importance_2016,riaud_influence_2017,chen_threedimensional_2018,ni_modelling_2019,das_acoustothermal_2019,kolesnik_periodic_2021,liu_threedimensional_2023}, while others have incorporated the Stokes drift term in the boundary treatment~\cite{vanneste_streaming_2011,nama_acoustic_2017,barnkob_acoustically_2018,bach_theory_2018,joergensen_theory_2021}. The no-slip condition assumes a zero Eulerian velocity~(the velocity field in a fixed spatial framework) at the wall, whereas the inclusion of Stokes drift enforces a zero Lagrangian velocity~(the motion of individual fluid particles). In bulk acoustic wave~(BAW) devices, the difference between these two boundary conditions has a negligible influence on the resulting streaming flow~\cite{nama_acoustic_2017}. For surface acoustic wave~(SAW) devices, however, the choice of boundary condition may lead to significantly different streaming patterns~\cite{nama_acoustic_2017}.

To evaluate the influence of the Stokes drift term on acoustic streaming patterns, this study compares simulation results obtained using the two types of second-order boundary conditions described above. To assess their accuracy, three-dimensional numerical results are compared with experimental measurements. Additionally, simplified modelling approaches are introduced to facilitate efficient three-dimensional simulations.

\section{Theories and methods}
\subsection{Model Setup}
\begin{figure}[htbp!]
 \centering
 \includegraphics[width=0.45\linewidth]{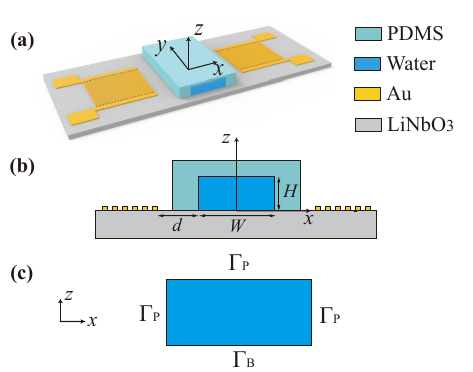}
 \caption{(a) Three-dimensional schematic and (b) two-dimensional $x$-$z$ cross-section of the SAW device. (c) Illustration of the fluid domain model used in the simulations.}
 \label{mesh_draft}
\end{figure}
The geometric model of a standard SAW device is illustrated in Fig.~\ref{mesh_draft}(a). The polydimethylsiloxane~(PDMS) chamber has an inner width $W$ and height $H$. The thickness of the top cover is on the millimetre scale, allowing reflections from the upper boundary to be neglected at the solid-liquid interface. The piezoelectric substrate consists of a 128$^\circ$ \textit{Y}-\textit{X} cut \ce{LiNbO3} wafer. Uniform interdigital transducers~(IDTs) are symmetrically distributed along the $y$-axis on both sides of the chamber, with an aperture $A$. The distance from the IDT front to the edge of the fluid domain is denoted as $d$.

To simplify the problem and focus on the essential physical aspects, both the two- and three-dimensional models in this study simulate only the fluid domain. Water at $T = 25^{\circ}\mathrm{C}$ is used as the working fluid. Numerical simulations are performed using COMSOL Multiphysics~\cite{comsol55}. Detailed material parameters are provided in Section C of the Supplemental Material ~\cite{supplementary}.

\subsection{Reynolds Stress Method~(RSM)}
The RSM method computes the steady-state acoustic streaming in acoustofluidic devices using perturbation theory~\cite{muller_numerical_2012}. Consider a fluid with initial pressure \(p_0\), initial temperature \(T_0\), and initial velocity \(\bm{v}_0 = \bm{0}\). The temperature \(T\), pressure \(p\), and velocity \(\bm{v}\) are expanded via the perturbation series $T-T_{0}  = T_{1}+T_{2}+\dots,\ 
    p-p_{0}  = p_{1}+p_{2}+\dots$, and
    $\bm{v}  = \bm{v}_{1} + \bm{v}_{2}+\dots,$
where the subscripts denote the order of expansion. Perturbation theory assumes \(p_2 \ll p_1 \ll p_0\) and \(T_2 \ll T_1 \ll T_0\). A low Mach number assumption is also applied, such that \(|\bm{v}_1|, |\bm{v}_2| \ll c_0\), where \(c_0\) is the isentropic speed of sound in the fluid, given by the state equation: \(p_1 = c_0^2 \rho_1\).

The first-order system comprises the energy conservation equation for the temperature perturbation \(T_1\), the continuity equation for the acoustic pressure \(p_1\), and the momentum equation for the velocity perturbation \(\bm{v}_1\),
\begin{equation}\label{eqn-rsm_1st}
  \begin{split}
    \frac{\partial T_1}{\partial t} &= \frac{\alpha_0 T_0}{\rho_0 C_p} \frac{\partial p_1}{\partial t} + \frac{k_{\mathrm{th}}}{\rho_0 C_p} \nabla^2 T_1, \\
    \frac{\gamma}{c_0^2} \frac{\partial p_1}{\partial t} &= \rho_0 \alpha_0 \frac{\partial T_1}{\partial t} - \nabla \cdot (\rho_0 \bm{v}_1), \\
    \rho_0 \frac{\partial \bm{v}_1}{\partial t} &= \nabla \cdot \left[ -\left(p_1 + \tilde{\mu} \nabla \cdot \bm{v}_1 \right) \bm{E} + \mu \left( \nabla \bm{v}_1 + (\nabla \bm{v}_1)^\mathrm{T} \right) \right].
  \end{split}
\end{equation}

Here, \(k_{\mathrm{th}}\) denotes the thermal conductivity, \(\alpha_0\) the thermal expansion coefficient, \(C_p\) the specific heat capacity, \(\gamma\) the adiabatic index, \(\mu\) the dynamic viscosity, and \(\mu_{\mathrm{B}}\) the bulk viscosity, with \(\tilde{\mu} = \frac{2}{3}\mu - \mu_{\mathrm{B}}\); \(\bm{E}\) represents the identity matrix. All first-order fields are assumed to be time-harmonic, i.e.,~they contain the harmonic factor \(e^{\mathrm{i}\omega t}\), where \(\omega\) is the angular frequency of SAWs, and $\mathrm{i}$ is the imaginary unit.

For most fluids, thermal effects in the first-order acoustic field are negligible. The second-order temperature \(T_2\) can therefore be decoupled from \(\bm{v}_2\) and \(p_2\)~\cite{muller_numerical_2012}. It should be noted that in acoustofluidics, the characteristic time scale of fluid particle motion is on the order of milliseconds, significantly longer than that of ultrasonic vibrations~(microseconds)~\cite{bruus_acoustofluidics_2012b}. Consequently, only the time-averaged components of the second-order fields are physically relevant. The time-averaged second-order momentum equation is
\begin{equation}
-\nabla \langle p_2 \rangle + \mu \beta \nabla (\nabla \cdot \langle \bm{v}_2 \rangle) + \mu \nabla^2 \langle \bm{v}_2 \rangle = -\bm{F}_{\mathrm{Reynolds}}, \label{eqn-decouple_2nd_NS}
\end{equation}
where \(\langle \cdot \rangle\) is the time-averaging operator and \(\beta = \mu_{\mathrm{B}} / \mu + 1/3\). The term \(\bm{F}_{\mathrm{Reynolds}}\) represents the Reynolds stress~(body force), defined as the divergence of the time-averaged acoustic momentum flux~\cite{lighthill_acoustic_1978},
\begin{subequations}
  \begin{align}
    \bm{F}_{\mathrm{Reynolds}} &= -\nabla \cdot \langle \rho_0 \bm{v}_1 \bm{v}_1 \rangle, \label{eqn-decouple_2nd_Reynolds1} \\
    &= -\rho_0 \langle \bm{v}_1 \cdot \nabla \bm{v}_1 \rangle - \left\langle \rho_1 \frac{\partial \bm{v}_1}{\partial t} \right\rangle. \label{eqn-decouple_2nd_Reynolds4}
  \end{align}
\end{subequations}
The transformation from Eq.~(\ref{eqn-decouple_2nd_Reynolds1}) to Eq.~(\ref{eqn-decouple_2nd_Reynolds4}) employs the first-order continuity equation. The time-averaged second-order continuity equation is
\begin{equation}
\rho_0 \nabla \cdot \langle \bm{v}_2 \rangle = -\nabla \cdot \langle \rho_1 \bm{v}_1 \rangle. \label{eqn-decouple_ce}
\end{equation}
Here, \(Q_{\mathrm{ms}} = -\nabla \cdot \langle \rho_1 \bm{v}_1 \rangle\) denotes the mass source term, representing the divergence of the time-averaged acoustic momentum flux. Together, Eqs.~(\ref{eqn-decouple_2nd_NS}) and~(\ref{eqn-decouple_ce}) form the governing system for the second-order steady streaming. The right-hand sides, \(\bm{F}_{\mathrm{Reynolds}}\) and \(Q_{\mathrm{ms}}\), serve as source terms for the momentum and continuity equations, respectively, driving the acoustic streaming~\cite{muller_numerical_2012}.

\subsection{First-order Boundary Conditions}
For the pure fluid domain model, the first-order acoustic boundary conditions consist of impedance and velocity boundaries. As illustrated in the two-dimensional model in Fig.~\ref{mesh_draft}(c), \(\Gamma_{\mathrm{P}}\) denotes the interface between the PDMS and the fluid, and \(\Gamma_{\mathrm{B}}\) the interface between the piezoelectric substrate and the fluid. The impedance boundary condition is prescribed as
\begin{equation}
    \bm{n} \cdot \nabla p_1 + \mathrm{i} \frac{\omega \rho_0}{\rho_\mathrm{P} c_{\mathrm{P}}} p_1 = 0, \quad \text{on } \Gamma_{\mathrm{P}},
\end{equation}
where \(\rho_\mathrm{P}\) and \(c_{\mathrm{P}}\) are the density and speed of sound in PDMS, respectively. Numerically, this condition is implemented as a Dirichlet boundary condition via a Lagrange multiplier~\cite{bruus_acoustofluidics_2012}.

On the fluid-solid interface \(\Gamma_{\mathrm{P}} \cup \Gamma_{\mathrm{B}}\), let \(\bm{\xi}_0\) be the initial position of a solid particle and \(\bm{\xi}(\bm{\xi}_0, t)\) its position at time~\(t\). The displacement is given by
\begin{equation}
\bm{\xi}(\bm{\xi}_0, t) = \bm{\xi}_0 + \bm{\xi}_1(\bm{\xi}_0) e^{\mathrm{i}\omega t}. \label{eqn-displacment}
\end{equation}
Owing to the viscous boundary layer, the fluid velocity \(\bm{v}(\bm{r}, t)\) at the wall position \(\bm{\xi}(\bm{\xi}_0, t)\) must equal the vibration velocity of the solid at that point, \(\bm{u}_1 e^{\mathrm{i}\omega t}\). Using perturbation expansion, this yields
\begin{equation}
\bm{v}_1(\bm{\xi}, t) + \bm{v}_2(\bm{\xi}, t) = \bm{u}_1(\bm{\xi}_0) e^{\mathrm{i}\omega t}. \label{eqn-true_no_slip}
\end{equation}
Expanding the left-hand side in a Taylor series around \(\bm{\xi}_0\) and retaining terms up to second order~\cite{bach_theory_2018} gives
\begin{equation}
\begin{split}
&\bm{v}_1(\bm{\xi}, t) + \bm{v}_2(\bm{\xi}, t) \\
&\approx \bm{v}_1(\bm{\xi}_0, t) + \bm{\xi}_1(\bm{\xi}_0) \cdot \nabla \bm{v}_1(\bm{\xi}_0) + \bm{v}_2(\bm{\xi}_0, t). \label{eqn-taylor_expansion}
\end{split}
\end{equation}
Matching terms by order, the first-order velocity boundary condition on \(\Gamma_{\mathrm{B}}\) becomes
\begin{equation}
\bm{v}_1(\bm{\xi}_0, t) = \bm{u}_1(\bm{\xi}_0) e^{\mathrm{i}\omega t}, \quad \text{on } \Gamma_{\mathrm{B}}. \label{eqn-no_slip_1st_order}
\end{equation}

For the two-dimensional~(\(x\)–\(z\)) model of a standing SAW~(SSAW) device, the vibration velocity \(\bm{u}_1 = \mathrm{i}\omega \bm{\xi}_1 = (u_x, u_z)\) is given by the superposition
\begin{equation}\label{eqn-SSAW-disp}
    \begin{aligned}
    u_x &= \mathrm{i} K \xi_1 \omega \left[ e^{(-\mathrm{i} k_s - \alpha)(x + \frac{W}{2})} + e^{\mathrm{i}\Delta\phi} e^{(\mathrm{i} k_s + \alpha)(x - \frac{W}{2})} \right], \\
    u_z &= -\xi_1 \omega \left[ -e^{(-\mathrm{i} k_s - \alpha)(x + \frac{W}{2})} + e^{\mathrm{i}\Delta\phi} e^{(\mathrm{i} k_s + \alpha)(x - \frac{W}{2})} \right],
    \end{aligned}
\end{equation}
where \(\xi_1\) is the amplitude of the first order displacement $\bm{\xi}_1$, \(\lambda_{\mathrm{s}}\) the SAW wavelength, \(k_{\mathrm{s}} = 2\pi/\lambda_{\mathrm{s}}\) the wavenumber, \(\alpha = \rho_0 c_0 / (\rho_{\mathrm{LN}} c_{\mathrm{LN}} \lambda_{\mathrm{s}})\) the attenuation factor of the leaky SAW, and \(\Delta\phi\) the phase difference between left and right travelling waves.

For a traveling SAW (TSAW) model in the \(x\)–\(z\) plane, assuming wave propagation in the \(+x\) direction, \(\bm{u}_1 = (u_x, u_z)\) retains only the first term in the brackets in Eq.~(\ref{eqn-SSAW-disp}). The ratio \(K = 0.86\) between the \(x\)- and \(z\)-components of the substrate displacement \(\bm{\xi}_1\) is taken from the curve-fitting results of Devendran~\textit{et al.}~\cite{devendran_importance_2016}.

\subsection{Second-Order Boundary Conditions}
Based on Eqs.~(\ref{eqn-true_no_slip}) and~(\ref{eqn-taylor_expansion}), the following relation exists,
\begin{equation}
\bm{v}_2(\bm{\xi}_0, t) = -\bm{\xi}_1(\bm{\xi}_0) \cdot \nabla \bm{v}_1(\bm{\xi}_0). \label{eqn-2nd_order_taylor}
\end{equation}
Hence, the second-order boundary condition becomes
\begin{equation}
\langle \bm{v}_2 \rangle = -\left\langle \frac{1}{\mathrm{i}\omega} \bm{v}_1 \cdot \nabla \bm{v}_1 \right\rangle = -\bm{v}^{\mathrm{SD}}, \quad \text{on } \Gamma_{\mathrm{P}} \cup \Gamma_{\mathrm{B}}. \label{eqn-stokes_drift}
\end{equation}
Here, \(\bm{v}^{\mathrm{SD}}\) denotes the Stokes drift velocity. This~(the Stokes slip) boundary condition, referred to as the SD condition in this work, is derived from the Taylor expansion truncated beyond second-order terms and assumes zero fluid particle velocity at the wall~\cite{vanneste_streaming_2011,nama_acoustic_2017,bach_theory_2018}.

Alternatively, some studies adopt a no-slip boundary condition by ignoring the Stokes drift term and enforcing zero Eulerian velocity at the wall~\cite{nyborg_acoustic_1958,lee_nearboundary_1989}, i.e.,
\begin{equation}
\langle \bm{v}_2 \rangle = \bm{0}, \quad \text{on } \Gamma_{\mathrm{P}} \cup \Gamma_{\mathrm{B}}. \label{eqn-no_slip}
\end{equation}
This is referred to as the NS condition hereafter. As noted in the Introduction, the NS condition has been widely used in acoustofluidic streaming simulations.

In RSM, the steady streaming is described by the Eulerian velocity \(\bm{v}^{\mathrm{E}}\), where \(\langle \bm{v}_2 \rangle = \bm{v}^{\mathrm{E}}\) represents the Eulerian streaming field. The Lagrangian velocity \(\bm{v}^{\mathrm{L}}\), which describes the actual motion of fluid particles, is given as
\begin{equation}
\bm{v}^{\mathrm{L}} = \bm{v}^{\mathrm{E}} + \bm{v}^{\mathrm{SD}}. \label{eqn-Lagrangian}
\end{equation}
Experimentally, particle image velocimetry~(PIV) measurements of acoustic streaming typically capture the Lagrangian velocity \(\bm{v}^{\mathrm{L}}\).

\subsection{The Mesh Setup and Geometric Parameters}
For models employing RSM, boundary layer meshing is necessary on all boundaries. A quadrilateral mapped mesh is used near the boundaries, while the bulk domain is discretized with a free quadrilateral mesh. A mesh convergence study was conducted prior to numerical comparisons, with details provided in the Supplemental Material~(Sections A \& B)~\cite{supplementary}. All subsequent two-dimensional simulations of the acoustofluidic chip adhere to the mesh settings determined by the convergence criteria.

Let \(\lambda_{\mathrm{f}}\) denote the wavelength of the longitudinal wave in the fluid and \(\theta_{\mathrm{R}} = \sin^{-1}(c_0 / c_{\mathrm{s}})\) the Rayleigh angle at which SAWs leak into the fluid from the substrate, where \(c_{\mathrm{s}}\) is the phase velocity of SAWs. The wavelength corresponding to the vertical component of the wave vector in the fluid is then given by \(\lambda_{\mathrm{v}} = \lambda_{\mathrm{f}} / \cos\theta_{\mathrm{R}}\), consistent with the definition by Kolesnik~\textit{et al.}~\cite{kolesnik_periodic_2021}. 

The channel width is defined as \(W = w \lambda_{\mathrm{s}}\), the height as \(H = h \lambda_{\mathrm{v}}\), and the viscous boundary layer thickness as \(\delta_{\mathrm{v}} = \sqrt{2\mu / (\omega \rho_0)}\), where \(w\) and \(h\) are dimensionless scaling parameters.

\subsection{Three-Dimensional Simulation}
To evaluate which boundary condition, NS or SD, yields results better matching the actual streaming pattern, three-dimensional simulation results are compared against experimental measurements. 

In the three-dimensional model, the velocity superposition formula incorporates a normalized displacement amplitude distribution \(\hat{\xi}(y)\) along the \(y\)-direction:
\begin{equation}\label{eqn-3d-disp}
\begin{aligned}
    u_x(\bm{r}) = u_x(x, z) \hat{\xi}(y), \quad
    u_z(\bm{r}) = u_z(x, z) \hat{\xi}(y),
\end{aligned}
\end{equation}
where \(\hat{\xi}(y)\) is computed using the angular spectrum method for anisotropic materials~\cite{tan_fast_1987} and the delay-and-sum of equivalent line sources~\cite{zhang_field_2021,ma_minimizing_2025a}. Detailed expressions are provided in the Supplemental Material~(Section G)~\cite{supplementary}.

Due to the high computational cost of solving full models employing the RSM method, a simplified modelling approach is adopted, which is validated on two-dimensional (\(x\)–\(z\)) models for the three devices considered in the experiments. The governing equations of simplified modelling and the validation results are included in the Supplemental Material~(Sections D \& E)~\cite{supplementary}. For simulations using the NS condition, the improved limiting velocity method~(ILVM) proposed by Chen~\textit{et al.}~\cite{chen_threedimensional_2018} is adopted, with the inclusion of an additional viscous acoustic body force~\cite{riaud_influence_2017, bach_theory_2018}. When applying the SD condition, an approach equivalent to that of Bach \& Bruus~\cite{bach_theory_2018} is followed, wherein an equivalent slip velocity is imposed on \(\Gamma_{\mathrm{B}}\), while the Stokes slip condition is retained on \(\Gamma_{\mathrm{P}}\). Free triangular and swept meshes are used in all three-dimensional simulations.

\subsection{Device Fabrication and Experimental Setup}
The IDTs were fabricated via photolithography and magnetron sputtering, depositing a Au/Cr layer (150/5~nm) onto a 2-inch-diameter, $500~\upmu\mathrm{m}$-thick 128$^\circ$ \textit{Y}-\textit{X} \ce{LiNbO3} wafer. The microfluidic chamber was manufactured using standard soft lithography and replica moulding with PDMS~(Sylgard 184, Dow Corning, Midland, MI, USA). The chamber was bonded to the substrate, aided by a plasma cleaner~(Pluto-T, Plutovac, Shanghai, China). The electrical properties of the IDTs were characterised with a vector network analyzer ~(VNA2180, ArraySolutions, Texas, USA) to identify the operating frequency range for the SAW device. Each IDT was driven by a signal generator~(33622A, Keysight, Colorado Springs, CA, USA) connected to a power amplifier~(ZHL-5W-1X+, Mini-Circuits, New York, USA).

In the experiments, $5~\upmu\mathrm{m}$-diameter polystyrene~(PS) spheres~(Baseline, Tianjin, China) were used to probe the acoustic pressure in the SSAW device based on acoustophoretic trajectory tracking. For PIV measurements of the acoustic streaming patterns, $500~\mathrm{nm}$ PS particles~(Zhongkekeyou, Beijing, China) served as tracers. The acoustofluidic device was mounted on an inverted microscope~(IX83, Olympus, Tokyo, Japan), and particle motion was captured using a high-speed camera~(FASTCAM Mini UX100, Photron, Tokyo, Japan). Acquired images were processed using ImageJ~(NIH, Bethesda, MD, USA) and MATLAB~(R2023b, MathWorks, Natick, MA, USA).

\section{RESULTS}
\begin{figure*}[htbp!]
 \centering
 \includegraphics[width=1\linewidth]{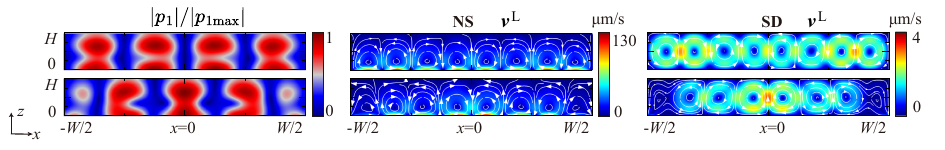}
 \caption{Distributions of the normalized absolute acoustic pressure $|p_{1}|/|p_{1\mathrm{max}}|$ and the Lagrangian velocity $\bm{v}^{\mathrm{L}}$ in the channel $x$-$z$ cross-section for the NS and SD boundary conditions. Results are shown for an SSAW device with parameters $h=0.75$, $w=2$, $\lambda_{\mathrm{s}}=200\ \upmu\mathrm{m}$, $\lambda_{\mathrm{v}}=80.7\ \upmu\mathrm{m}$, and vibration amplitude $\xi_{1}=0.3\ \mathrm{nm}$. The first row corresponds to a phase difference $\Delta\phi = 0$; the second row to $\Delta\phi = \pi$. Streamlines depict the flow field; arrows indicate local velocity direction.}
 \label{model_difference_comparison}
\end{figure*}

\subsection{Impact of SD Boundary in SSAW Devices\label{results.1}}
For the SSAW device with wavelength $\lambda_{\mathrm{s}} = 200~\upmu\mathrm{m}$, channel width $W = 2\lambda_{\mathrm{s}}$, and height $H = 0.75\lambda_{\mathrm{v}}$, the normalized acoustic pressure field and the Lagrangian velocity $\bm{v}^{\mathrm{L}}$ distribution, computed using both the NS and SD conditions, are shown in Fig.~\ref{model_difference_comparison} for phase differences $\Delta\phi = 0$ and $\Delta\phi = \pi$.

When $\Delta\phi = 0$, an acoustic pressure node sits at $x = 0$, and four antinodes are present along the horizontal direction. In contrast, $\Delta\phi = \pi$ induces an antinode at $x = 0$, and only three antinodes are observed. These differences result in distinct streaming patterns.

\begin{figure}[htbp!]
 \centering
 \includegraphics[width=0.45\linewidth]{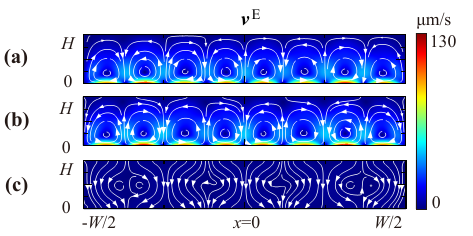}
 \caption{Eulerian velocity $\bm{v}^{\mathrm{E}}$ distributions for the SSAW device ($\Delta\phi=0$) under different modeling approaches: (a) NS condition, (b) SD condition with body force $\bm{F}_{\mathrm{Reynolds}} = 0$, and (c) full SD condition. Streamlines depict the flow field; arrows indicate local velocity direction. Device parameters are identical to those in Fig.~\ref{model_difference_comparison}.}
 \label{Stokes_drift_decompose}
\end{figure}

For the NS condition, the $\bm{v}^{\mathrm{L}}$ distribution exhibits eight  vortices across the two-wavelength channel width in both cases, though the vortex directions are opposite. At pressure antinodes, $\bm{v}^{\mathrm{L}}$ consistently points in the $-z$ direction, opposite to the acoustic intensity direction~($+z$). The distributions of $\bm{v}^{\mathrm{E}}$ and $\bm{v}^{\mathrm{L}}$ are nearly identical. For example, at $\Delta\phi = 0$, the Eulerian velocity $\bm{v}^{\mathrm{E}}$ (Fig.~\ref{Stokes_drift_decompose}(a)) closely matches $\bm{v}^{\mathrm{L}}$, except that $\bm{v}^{\mathrm{L}}$ streamlines are not fully closed near the wall, resulting in a net flux through the domain, whereas $\bm{v}^{\mathrm{E}}$ exhibits zero net flux.

Under the SD condition, the $\bm{v}^{\mathrm{L}}$ field shows eight vortices for $\Delta\phi = 0$ but only six for $\Delta\phi = \pi$. At antinodes, $\bm{v}^{\mathrm{L}}$ aligns with the acoustic intensity direction~($+z$). The vortex directions under the SD condition are opposite to those under NS. Vortex centers in the SD case are located higher than in the NS case, and the streaming velocity amplitude is 1–2 orders of magnitude smaller. Here, a significant discrepancy arises between $\bm{v}^{\mathrm{E}}$ and $\bm{v}^{\mathrm{L}}$. For $\Delta\phi = 0$, the Eulerian field $\bm{v}^{\mathrm{E}}$ (Fig.~\ref{Stokes_drift_decompose}(c)) shows almost no vortical motion in the bulk and is confined primarily  within the bottom boundary layer.

\begin{figure}[htbp!]
 \centering
 \includegraphics[width=0.45\linewidth]{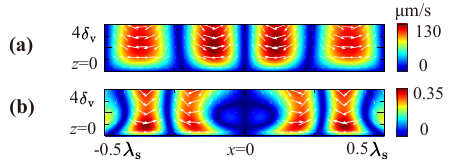}
 \caption{Velocity distributions in an SSAW device, obtained near the bottom boundary $\Gamma_{\mathrm{B}}$ within the region $-\lambda_{\mathrm{s}}/2 < x < \lambda_{\mathrm{s}}/2$, $0 < z < 4\delta_\mathrm{v}$: (a) $\bm{v}^{\mathrm{E}}$ under the NS condition and (b) $\bm{v}^{\mathrm{L}}$ under the SD condition. White arrows indicate flow direction. Device parameters match Fig.~\ref{model_difference_comparison} ($\Delta\phi = 0$).}
 \label{Inner_layer_SSAW}
\end{figure}

Near the bottom boundary $\Gamma_{\mathrm{B}}$, the acoustic streaming distributions exhibit both similarities and differences under the two boundary conditions. For the case $\Delta\phi = 0$ (Fig.~\ref{Inner_layer_SSAW}), when the NS condition is applied, $\bm{v}^{\mathrm{E}}$ displays a velocity gradient in the $z$-direction, with a local maximum near $z = 2\delta_\mathrm{v}$. The outer flow is primarily driven by the velocity gradient within the boundary layer, rather than by the inner-boundary-layer vortex~(Schlichting streaming) as observed in BAW devices~\cite{muller_numerical_2012}. This behaviour aligns with the results of Devendran~\textit{et al.}~\cite{devendran_importance_2016}, who also employed the NS condition in SAW streaming simulations.

When employing the SD condition, $\bm{v}^{\mathrm{L}}$ near $\Gamma_{\mathrm{B}}$ also exhibits a $z$-direction velocity gradient, peaking around $z = \delta_\mathrm{v}$. Similarly, the outer streaming is driven by the inner velocity gradient, consistent with the NS case. However, the direction of $\bm{v}^{\mathrm{L}}$ in the SD case is essentially opposite to that of $\bm{v}^{\mathrm{E}}$ in the NS case.

To elucidate the differences in streaming patterns between the two boundary conditions, consider the case $\Delta\phi = 0$ with the SD condition and body force $\bm{F}_{\mathrm{Reynolds}} = 0$. The resulting Eulerian velocity $\bm{v}^{\mathrm{E}}$, shown in Fig.~\ref{Stokes_drift_decompose}(b), is induced solely by the vibrating wall. Its amplitude is comparable to that under the NS condition~(Fig.~\ref{Stokes_drift_decompose}(a)), but the vortex directions are opposite. By the superposition principle for linear systems, the total $\bm{v}^{\mathrm{E}}$ under the full SD condition can be viewed as the sum of the NS-case $\bm{v}^{\mathrm{E}}$ and the field from Fig.~\ref{Stokes_drift_decompose}(b). This superposition leads to near cancellation of the Eulerian vortices, as seen in Fig.~\ref{Stokes_drift_decompose}(c).

Thus, the competition between the Stokes slip boundary and the Reynolds stress body force $\bm{F}_{\mathrm{Reynolds}}$ results in a Lagrangian streaming field $\bm{v}^{\mathrm{L}}$ under the SD condition that has opposite rotation and a significantly smaller magnitude compared to the NS condition.

\begin{figure*}[htbp!]
 \centering
 \includegraphics[width=0.95\linewidth]{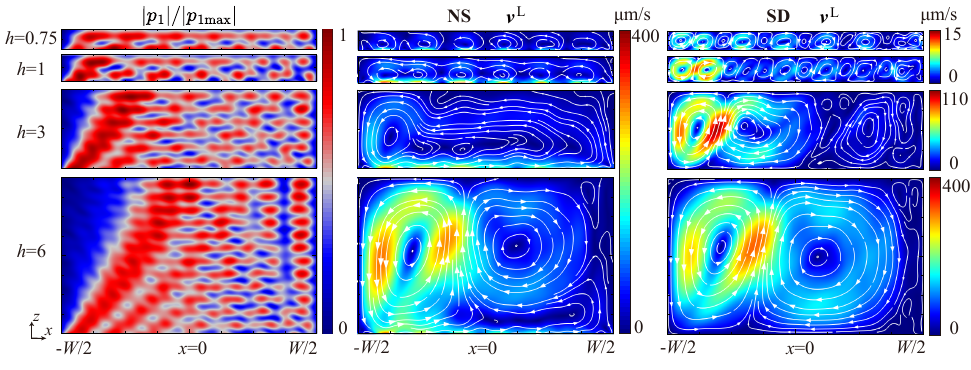}
 \caption{Normalized acoustic pressure $|p_1|/|p_{1\mathrm{max}}|$ and Lagrangian velocity $\bm{v}^{\mathrm{L}}$ distributions for TSAW devices with different channel heights $h$, under NS and SD conditions. Wave propagation is along $+x$. Streamlines depict the flow field; arrows indicate local velocity direction. Parameters:  $w=4$, $\lambda_{\mathrm{s}}=200\ \upmu\mathrm{m}$, $\lambda_{\mathrm{v}}=80.7\ \upmu\mathrm{m}$, $\xi_1=1\ \mathrm{nm}$.}
 \label{NS&SD_TSAW_comparison}
\end{figure*}

\subsection{Impact of SD Boundary in TSAW Devices\label{results.2}}
For the TSAW device with $\lambda_{\mathrm{s}} = 200~\upmu\mathrm{m}$ and channel width $W = 4\lambda_{\mathrm{s}}$, a parametric study was conducted over the normalized channel height $h = H/\lambda_{\mathrm{v}}$ ($h = 0.75, 1, 3, 6$). The resulting normalized acoustic pressure and Lagrangian velocity $\bm{v}^{\mathrm{L}}$ distributions, simulated under both NS and SD conditions, are shown in Fig.~\ref{NS&SD_TSAW_comparison}.

First consider the NS condition. At low channel heights~($h = 0.75, 1$), six streaming vortices with identical rotational direction are observed in the $x$–$z$ cross-section, all attributed to boundary-driven flow. The vortex spacing approximately follows $\lambda_{\mathrm{R}} = \lambda_{\mathrm{f}}/(1 - c_{\mathrm{f}}/c_{\mathrm{s}})$~\cite{devendran_huygensfresnel_2017,kolesnik_periodic_2021}, consistent with diffraction effects within the channel. At $h = 3$, boundary-driven streaming remains pronounced near the bottom wall. In the bulk region, where the traveling wave impinges at the Rayleigh angle $\theta_{\mathrm{R}}$, an Eckart streaming vortex emerges due to non-uniform acoustic intensity, whose size is comparable to $H$. The bulk vortices maintain a consistent counterclockwise rotation. At $h = 6$, the Eckart streaming near the anechoic corner evolves into the so-called ``lobe streaming''~\cite{fakhfouri_surface_2018}, exhibiting a rotation direction opposite to that of its adjacent vortices. Boundary-driven streaming near the bottom wall remains clearly visible.

These findings align with those of Kolesnik~\textit{et al.}~\cite{kolesnik_periodic_2021} regarding the height-dependent interplay between periodic boundary-driven streaming and Eckart streaming in TSAW devices, despite differences in channel width and the use of Eulerian velocity $\bm{v}^{\mathrm{E}}$ in their study.

Distinct differences emerge under the SD condition. At $h = 0.75$, a lobe-streaming vortex appears with opposite rotation to its neighbors. Except for this counterclockwise lobe vortex, the overall flow in the channel is predominantly clockwise. At $h = 1$, the lobe vortex is stronger and rotates oppositely to adjacent vortices; no unified flow direction is observed across the channel. For greater heights ($h = 3, 6$), the near-wall amplitude is markedly reduced. The lobe vortex persists with opposite rotation relative to its neighbors; aside from this, the remaining vortices show generally consistent rotation.

These differences arise from the influence of the Stokes drift term. Under the SD condition, competition between the Stokes slip boundary and $\bm{F}_{\mathrm{Reynolds}}$ substantially reduces the amplitude of the boundary-driven Lagrangian flow $\bm{v}^{\mathrm{L}}$, making the attenuation-driven Eckart streaming more dominant. Consequently, lobe streaming emerges even at relatively small channel heights in the SD case, whereas it only becomes distinct at larger $h$ under the NS condition. However, when the channel height satisfies $H > W \cot \theta_{\mathrm{R}}$, Eckart flow is expected to dominate in both cases, and the results obtained with NS and SD conditions should converge, with lobe streaming vortices prevailing throughout the channel.

\begin{figure*}[!ht]
 \centering
 \includegraphics[width=0.95\linewidth]{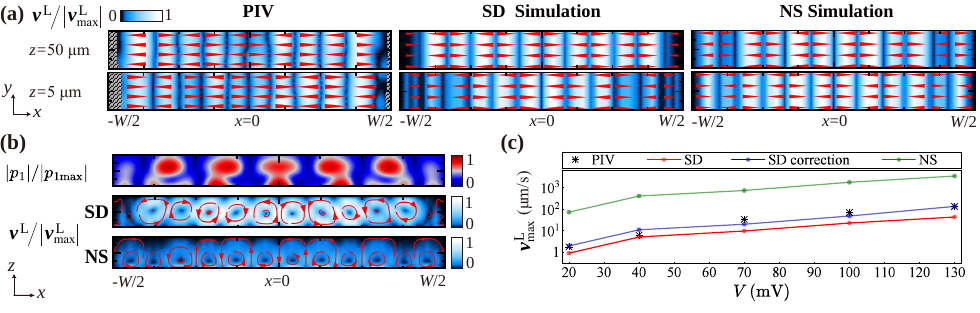}
 \caption{Streaming fields in Device A. (a)~$x$-component of velocity in the $x$-$y$ plane at $z=50\ \upmu\mathrm{m}$ and $z=5\ \upmu\mathrm{m}$ from experiment~(PIV) and simulations~(SD/NS conditions). The $y$-coordinates range from $-40\ \upmu\mathrm{m}$ to $+40\ \upmu\mathrm{m}$. (b)~Normalized acoustic pressure and Lagrangian velocity $\bm{v}^{\mathrm{L}}$ in the $x$-$z$ cross-section ($y=0$). (c)~Maximum measured and simulated values of $v^{\mathrm{L}}_x$ vs. driving voltage $V$. Parameters: $H_{\mathrm{A}}=56\ \upmu\mathrm{m}$, $\Delta\phi=\pi$. Color maps show normalized amplitude; arrows indicate vector fields.}
 \label{Boundary_comparison}
\end{figure*}

\subsection{Comparing the Measured and Simulated Streaming Fields in SSAW Devices}
For an SSAW device, PIV was performed to characterise the streaming patterns in the fluid domain corresponding to the central region of the IDT aperture~(see the experimental recording in the Supplementary Video S1). A comparison between experimental and numerical results in terms of vortex patterns and velocity magnitudes is presented in Fig.~\ref{Boundary_comparison}. The parameters of the SSAW device~(Device A) are listed in Table~\ref{table 2}, where $A$ denotes the aperture width and $N$ the number of finger pairs.

\begin{table}[h]
	\centering
	\caption{Parameters of the three devices used in experiments, all of which have channel widths of $W=600\ \upmu\mathrm{m}$. }
\begin{tabular}{|c|c|c|c|c|c|c|}
\hline
\diagbox{Device}{Parameters} & $\lambda_{\mathrm{s}}$ & $H$ & $A$ & $d$ & $N$\\
\hline
Device A & $200\ \upmu\mathrm{m}$ & $56\ \upmu\mathrm{m}$ & $4\ \mathrm{mm}$ & $1.6\ \mathrm{mm}$ & 20 \\
\hline
Device B & $80\ \upmu\mathrm{m}$ & $62\ \upmu\mathrm{m}$ & $4\ \mathrm{mm}$ & $1.25\ \mathrm{mm}$ & 30 \\   
\hline
Device C & $80\ \upmu\mathrm{m}$ & $131\ \upmu\mathrm{m}$ & $4\ \mathrm{mm}$ & $1.25\ \mathrm{mm}$ & 30 \\
\hline
\end{tabular}
	\label{table 2}
\end{table}

PS particles of $500~\mathrm{nm}$-diameter were introduced into the channel at a concentration of $5\ \mathrm{mg/mL}$. The motion of particles was recorded at a frame rate of 500~fps immediately after acoustic activation and before significant particle aggregation occurred. Acoustic streaming velocities in the $x$–$y$ plane at heights close to channel ceiling and floor~($z=5\ \upmu\mathrm{m}$ and $z=50\ \upmu\mathrm{m}$) were subsequently extracted using the PIVLab package in MATLAB~\cite{PIVlab}.

After $5~\upmu\mathrm{m}$ PS particles were introduced at a concentration of $1\ \mathrm{mg/mL}$, their trajectories were tracked using ImageJ. Then, the acoustic pressure amplitude~($\left|p_1\right|$) was determined via the particle tracking method described by Barnkob~\textit{et al.}~\cite{barnkob_measuring_2010,liu_lookup_2024}, with $\left|p_1\right|$ at different driving voltages obtained through curve fitting. From these values, the displacement amplitude $\xi_1$ was derived, enabling the calibration of streaming field amplitudes in numerical simulations under both boundary conditions. Further details are provided in the Supplemental Material~(Section I)~\cite{supplementary}.

First, the vortex patterns from PIV and simulations are compared. As shown in Fig.~\ref{Boundary_comparison}(a), when the channel center is a pressure antinode~($\Delta\phi = \pi$), ten distinct streaming vortices with alternating rotation are observed experimentally. The maximum velocity at $z = 5~\upmu\mathrm{m}$ exceeds that at $z = 50~\upmu\mathrm{m}$ for all voltages. Higher voltages accelerate the formation of boundary-layer vortices. The in-plane velocity component $\bm{v}^{\mathrm{L}}_x$ from SD-condition simulations matches the experimental observations in both direction and number of vortices. Slight discrepancies occur only near the sidewalls. In contrast, NS-condition simulations yield $\bm{v}^{\mathrm{L}}_x$ directions opposite to the experimental results and predict twelve vortices. When the phase is adjusted so that a pressure node lies at the channel center~($\Delta\phi = 0$), the experimental vortex directions remain consistent with SD-condition simulations and opposite to NS-condition results. Additional details are included in the Supplemental Material~(Fig.~S14)~\cite{supplementary}.

\begin{figure*}[htbp!]
 \centering
 \includegraphics[width=0.95\linewidth]{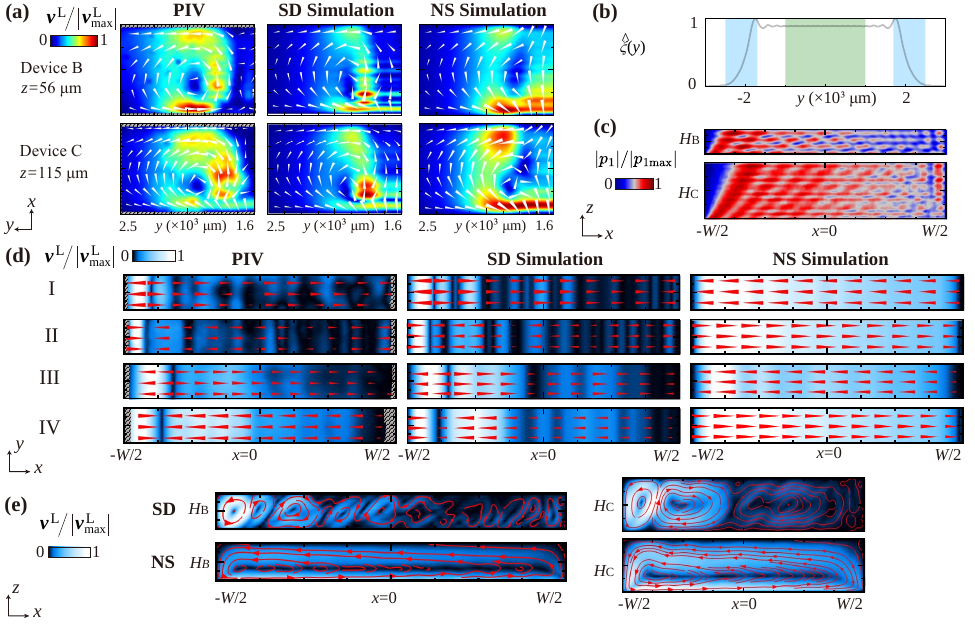}
 \caption{Peripheral and lobe streaming in Devices B \& C. (a) Peripheral streaming in the $x$-$y$ plane at measured $z$ heights from PIV and simulations. The $x$-coordinate ranges from $-300\ \upmu\mathrm{m}$ to $+300\ \upmu\mathrm{m}$. (b) Normalized displacement amplitude $\hat{\xi}(y)$; blue and green regions indicate measurement zones for peripheral and lobe streaming, respectively. (c) Normalized acoustic pressure in the $x$-$z$ plane ($y=0$). (d) Lobe streaming distributions in the $x$-$y$ plane at different $z$. The $y$-coordinate ranges from $-40\ \upmu\mathrm{m}$ to $+40\ \upmu\mathrm{m}$. For Device B: I \& II correspond to $z=56\ \upmu\mathrm{m}$ and $z=5\ \upmu\mathrm{m}$. For Device C: III \& IV correspond to $z=115\ \upmu\mathrm{m}$ and $z=10\ \upmu\mathrm{m}$. (e) Simulated Lagrangian velocity $\bm{v}^{\mathrm{L}}$ in the $x$-$z$ plane ($y=0$) for SD and NS conditions. Device parameters: $H_{\mathrm{B}}=62\ \upmu\mathrm{m}$, $H_{\mathrm{C}}=131\ \upmu\mathrm{m}$. Color maps: normalized amplitude; arrows: vector fields.}
 \label{Peri&Lobe_comparison}
\end{figure*}

Second, the velocity magnitudes are compared. Figure~\ref{Boundary_comparison}(c) shows the maximum values of $v^{\mathrm{L}}_x$ from PIV and simulations across driving voltages. NS-condition results overestimate the experimental values by 1–2 orders of magnitude. Note that the thermoviscous correction barely changes the magnitude in the NS case. SD-condition results without thermoviscous correction bracket the PIV data at $V = 20$ and $40~\mathrm{mV}$, while at higher voltages ($V = 70$, $100$, and $130~\mathrm{mV}$), the results incorporating thermoviscous correction~(corresponding to ``SD correction'' in Fig.~\ref{Boundary_comparison}(c)) agree closely with experiments. Section H in the Supplemental Material shows details of the thermoviscous correction~\cite{supplementary}.

Thus, for the SSAW device having~$h \sim 0.7$ (Device A), numerical simulations employing the SD condition reproduce the experimentally observed streaming patterns more accurately than those using the NS condition.

\subsection{Experimental Validation of Streaming Patterns in TSAW Devices}
For TSAW devices~(B and C in Table~\ref{table 2}), PIV results are taken to approximate the Lagrangian velocity $\bm{v}^{\mathrm{L}}$~(see the experimental recording in the supplementary Video S2). Measurements were first performed in the peripheral flow region near the IDT aperture edges ($1.6 \times 10^{3}~\upmu\mathrm{m} < y < 2.5 \times 10^{3}~\upmu\mathrm{m}$, corresponding to the left blue region of Fig.~\ref{Peri&Lobe_comparison}(b)), at heights close to the channel ceilings~(Device B : $z=56\ \upmu\mathrm{m}$. Device C : $z=115\ \upmu\mathrm{m}$). 

Comparisons with simulations are shown in Fig.~\ref{Peri&Lobe_comparison}(a). Numerical results using the SD condition accurately reproduce the size and center positions of the peripheral streaming vortices observed experimentally. In contrast, simulations using the NS condition fail to capture the vortex structure for Device B at $z = 56~\upmu\mathrm{m}$, showing a uniformly negative $x$-component of velocity. For Device C, a vortex is present at $z = 115~\upmu\mathrm{m}$, but the velocity direction near $y = 1.6 \times 10^{3}~\upmu\mathrm{m}$ is opposite to the experimental observation.

The normalized amplitude distribution $\hat{\xi}(y)$~(blue region in Fig.~\ref{Peri&Lobe_comparison}(b)) indicates a strong acoustic field gradient near the aperture edge, suggesting that peripheral vortices in high-frequency TSAW devices arise primarily from acoustic field inhomogeneity~\cite{collins_highly_2016a}, consistent with Eckart streaming. The Stokes drift term predominantly influences the vortex near $y = 1.6 \times 10^{3}~\upmu\mathrm{m}$. Competition between the Reynolds stress and the Stokes slip boundary suppresses the formation of a unidirectional channel-spanning flow~\cite{kolesnik_periodic_2021} in the $x$–$z$ plane. In NS-condition simulations, due to the absence of Stokes slip boundary, the Eckart streaming~(directed along $+x$, aligned with acoustic intensity) is much weaker than boundary-driven streaming~(along $-x$) near $y = 1.6 \times 10^{3}~\upmu\mathrm{m}$, leading to results inconsistent with experiments.

PIV measurements were also conducted in the central aperture region at heights close to channel bottoms and channel tops for both devices, with the results shown in Fig.~\ref{Peri&Lobe_comparison}(d). Experimentally, both devices exhibit distinct lobe vortices rotating opposite to their neighbors. The lobe size is comparable to the anechoic corner dimensions, with height $H$ and top width $\approx H \tan \theta_{\mathrm{R}}$. For Device B, velocities in regions far from the TSAW incidence align with the lobe direction but are much weaker. For Device C, the lobe vortex opposes the rest of the flow field at all measured heights.

The SD condition simulations generally match the experimental flow directions at corresponding $z$, with minor discrepancies in regions far from the incidence for Device B. In contrast, NS-condition simulations produce uniformly directed vortices throughout the channel, inconsistent with experiments. Normalized velocity distributions in the $x$–$z$ cross-section at $y = 0$ from 3D simulations are shown in Fig.~\ref{Peri&Lobe_comparison}(e), which align with the two-dimensional results in Fig.~\ref{NS&SD_TSAW_comparison} despite differences in operating frequency.

Therefore, for TSAW devices with $h \sim 1.9$ (Device B) and $h \sim 4$ (Device C), simulations employing the SD condition reproduce experimentally observed streaming patterns more accurately than the NS-condition cases.

\section{DISCUSSION}
\subsection{Physical Mechanism of Stokes Drift in Acoustic Streaming}
\begin{figure}[htbp!]
 \centering
 \includegraphics[width=0.45\linewidth]{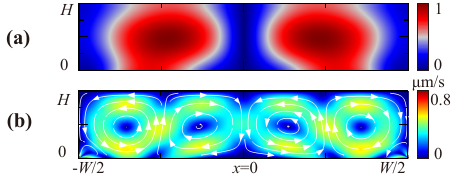}
 \caption{Distributions of (a) normalized acoustic pressure $|p_1|/|p_{1\mathrm{max}}|$ and (b) Lagrangian velocity $\bm{v}^{\mathrm{L}}$ under the SD condition. SSAW device parameters: $w=1$, $h=0.5$, $\lambda_{\mathrm{s}}=200\ \upmu\mathrm{m}$, $\lambda_{\mathrm{v}}=80.7\ \upmu\mathrm{m}$, $\xi_1=0.3\ \mathrm{nm}$, $\Delta\phi=0$.}
 \label{SSAW_discussion}
\end{figure}

\begin{figure}[htbp!]
 \centering
 \includegraphics[width=0.45\linewidth]{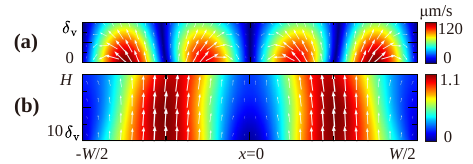}
 \caption{Stokes drift velocity $\bm{v}^{\mathrm{SD}}$ distribution (a) near ($0<z<\delta_\mathrm{v}$) and (b) far ($10\delta_\mathrm{v}<z<H$) from the bottom boundary. SSAW device parameters match Fig.~\ref{SSAW_discussion}.}
 \label{StokesDrift_Distribution}
\end{figure}

As shown in Section~\ref{results.1}, the Lagrangian velocity $\bm{v}^{\mathrm{L}}$ and Eulerian velocity $\bm{v}^{\mathrm{E}}$ differ significantly in SSAW device simulations when the SD condition is applied. It is therefore essential to analyze the distribution of the Stokes drift velocity $\bm{v}^{\mathrm{SD}}$. Taking the SSAW device in Fig.~\ref{SSAW_discussion} as an example, the spatial distribution of $\bm{v}^{\mathrm{SD}}$ under the SD condition is shown in Fig.~\ref{StokesDrift_Distribution}.

\begin{figure}[htbp!]
\centering
\includegraphics[width=0.45\linewidth]{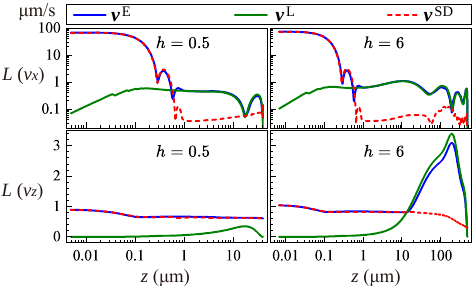}
\caption{Height profiles of the $x$-averaged $x$- and $z$-components of $|\bm{v}^{\mathrm{E}}|$, $|\bm{v}^{\mathrm{L}}|$, and $|\bm{v}^{\mathrm{SD}}|$ for two SSAW devices with different channel heights. Parameters: $w=1$, $\lambda_{\mathrm{s}}=200\ \upmu\mathrm{m}$, $\lambda_{\mathrm{v}}=80.7\ \upmu\mathrm{m}$, $\xi_1=0.3\ \mathrm{nm}$.}
\label{Stokes_drift_LProjection}
\end{figure}

Within the viscous bottom boundary layer, the Stokes drift velocity $\bm{v}^{\mathrm{SD}}$ is dominated by its $x$-component. Outside this layer, $\bm{v}^{\mathrm{SD}}$ aligns primarily with the acoustic intensity $\bm{I}$ in the bulk region and is dominated by its $z$-component. In the fluid domain of an SSAW device, the acoustic field in the $x$-direction is a partial standing wave, while the $z$-component is primarily a travelling wave due to acoustic absorption and low reflection at the PDMS interface. Only the travelling wave component contributes to the acoustic intensity $\bm{I}$. Neglecting viscosity, the Stokes drift velocity for a plane traveling wave satisfies~(see Section J in the Supplemental Material~\cite{supplementary}) $\bm{v}^{\mathrm{SD}} \approx \frac{1}{2\rho_{0}c_{0}^{2}}\bm{I}$, indicating that away from boundaries, $\bm{v}^{\mathrm{SD}}$ is approximately proportional to $\bm{I}$.

To compare $\bm{v}^{\mathrm{SD}}$ distributions across different channel heights, an averaging operator $L\left(\cdot\right) = \int_{0}^{W} \left| \cdot \right| \mathrm{d}x / W$ is defined,
and the height-dependent profiles of the $x$- and $z$-components of $|\bm{v}^{\mathrm{E}}|$, $|\bm{v}^{\mathrm{L}}|$, and $|\bm{v}^{\mathrm{SD}}|$ are analyzed, as shown in Fig.~\ref{Stokes_drift_LProjection}.

For both channel heights~($h=0.5, 6$), $\bm{v}^{\mathrm{SD}}$ behaves similarly near the bottom boundary. The $x$-component decays by nearly three orders of magnitude within $z = 0.1$–$1~\upmu\mathrm{m}$, while the $z$-component decays only slightly. Within the boundary layer, $|\bm{v}^{\mathrm{SD}}|$ and $|\bm{v}^{\mathrm{E}}|$ are comparable in magnitude but opposite in direction, resulting in a much smaller $|\bm{v}^{\mathrm{L}}|$. Outside the boundary layer, cancellation between $v^{\mathrm{SD}}_z$ and $v^{\mathrm{E}}_z$ leads to a $x$-averaged $|v^{\mathrm{L}}_z|$ smaller than $|v^{\mathrm{E}}_z|$ in the lower channel~($h=0.5$). In the higher channel~($h=6$), the effect of acoustic attenuation makes $|v^{\mathrm{L}}_z|$ and $|v^{\mathrm{E}}_z|$ comparable at $z > 10~\upmu\mathrm{m}$. For the channel with $h = 6$, $\bm{v}^{\mathrm{E}}$ and $\bm{v}^{\mathrm{L}}$ are similar in the bulk but differ significantly near the bottom boundary~(see Fig.~S15 in the Supplemental Material~\cite{supplementary}).

Thus, in SSAW devices, the Stokes drift term primarily influences near-wall boundary-driven streaming. Its effect diminishes with increasing channel height. When $H / \lambda_{\mathrm{v}} \gg 1$, acoustic attenuation dominates over boundary effects in the bulk region, and $\bm{v}^{\mathrm{E}}$ and $\bm{v}^{\mathrm{L}}$ converge under the SD condition. 

In TSAW devices, shear within the bottom boundary layer also causes $\bm{v}^{\mathrm{SD}}$ to be dominated by its $x$-component. Outside the boundary, $\bm{v}^{\mathrm{SD}}$ aligns with the oblique incident wave direction. The analysis follows a similar approach; details are provided in Fig.~S16 in the Supplemental Material~\cite{supplementary}.

\subsection{Comparison with Navier Slip Boundary Condition}
\begin{figure}[htbp!]
 \centering
 \includegraphics[width=0.45\linewidth]{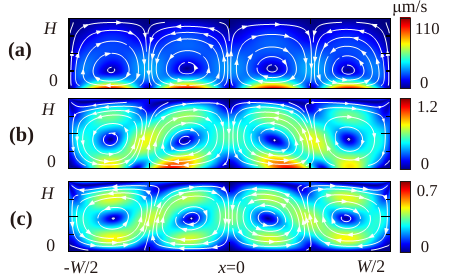}
 \caption{Second-order velocity $\langle\bm{v}_{2}\rangle$ from the Navier-slip method with varying slip lengths $L_\mathrm{s}$: (a) $L_\mathrm{s}=0.01\ \upmu\mathrm{m}$, (b) $L_\mathrm{s}=5\ \upmu\mathrm{m}$, (c) $L_\mathrm{s} \rightarrow \infty$ (free slip). SSAW device parameters match Fig.~\ref{SSAW_discussion}.}
 \label{methods_NavierSlip}
\end{figure}

To account for shear stress within the boundary layer and ensure the fluid responds to high-frequency tangential wall vibrations, Sachs~\textit{et al.} introduced a Navier slip model~\cite{sachs_acoustically_2022,lauga_microfluidics_2007}. By tuning the slip length~\(L_\mathrm{s} = r \delta_\mathrm{v}\), where \(r\) is a dimensionless parameter determined experimentally~\cite{xie_boundary_2014}, the resulting streaming patterns can be matched to measurements. In this approach, the first-order field incorporates Navier slip through an additional boundary stress and slip velocity. All other equations and boundary conditions remain consistent with the standard RSM method. Here, Navier slip is applied in modelling the SSAW device used in Fig.~\ref{SSAW_discussion}, with the results shown in Fig.~\ref{methods_NavierSlip}.

As \(L_\mathrm{s} \rightarrow 0\), the streaming pattern converges to that obtained with the NS condition; as \(L_\mathrm{s} \rightarrow \infty\), it approaches the free-slip~(or perfect-slip~\cite{lauga_microfluidics_2007}) limit~\cite{sachs_acoustically_2022}. For SSAW devices of fixed geometry, higher driving frequencies require smaller slip lengths to match experimental results~\cite{sachs_acoustically_2022}. 

For this device, the streaming pattern at \(L_\mathrm{s} = 0.01~\upmu\mathrm{m}\) agrees with the Lagrangian velocity \(\bm{v}^{\mathrm{L}}\) from RSM under the NS condition. In contrast, the result from RSM with the SD condition resembles those obtained with \(L_\mathrm{s} = 5~\upmu\mathrm{m}\) or \(L_\mathrm{s} \rightarrow \infty\). In the free-slip limit (\(L_\mathrm{s} \rightarrow \infty\)), boundary-driven streaming is negligible, and the flow is dominated by Eckart streaming~(interior streaming)~\cite{xie_boundary_2014}. Thus, the Stokes slip condition in RSM yields streaming patterns similar to the Navier slip method with free slip, differing only near the bottom boundary layer.

\subsection{Numerical Considerations and Alternative Formulations}
When using the RSM method, the mesh convergence of the second-order velocity $\langle\bm{v}_{2}\rangle$ is less robust under the SD condition than under the NS condition. Achieving a similar level of convergence requires a denser mesh, primarily due to the need to resolve second-order wall vibrations. Since the Stokes drift term originates from a second-order Taylor expansion that neglects higher-order terms, the Lagrangian velocity $\bm{v}^{\mathrm{L}}$, derived from the Eulerian velocity $\bm{v}^{\mathrm{E}}$, inevitably incurs some numerical error.

\begin{figure}[htbp!]
 \centering
 \includegraphics[width=0.45\linewidth]{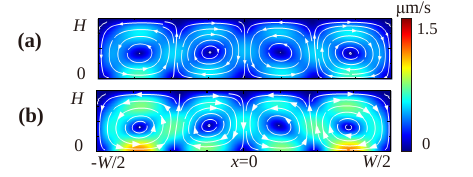}
 \caption{Lagrangian velocity $\bm{v}^{\mathrm{L}}$ from the ALE method (a) without and (b) with thermoviscous correction. SSAW device parameters match Fig.~\ref{SSAW_discussion}.}
 \label{methods_ALE}
\end{figure}

To circumvent the challenges associated with the Stokes slip boundary condition, the arbitrary Lagrangian–Eulerian~(ALE) method proposed by Nama~\textit{et al.} can be employed~\cite{nama_acoustic_2017,barnkob_acoustically_2018}. This method uses $\bm{v}^{\mathrm{L}}$ as the independent variable in the second-order system and imposes a no-slip condition $\bm{v}^{\mathrm{L}} = 0$ on $\Gamma_{\mathrm{P}} \cup \Gamma_{\mathrm{B}}$. The governing equations are Eq.~(S17) and Eq.~(S18) in the Supplemental Material~\cite{supplementary}, and the numerical implementation is carried out via the ``Weak Form'' module in COMSOL. Results for the SSAW device used in Fig.~\ref{SSAW_discussion}, obtained using the ALE method with and without thermoviscous correction, are shown in Fig.~\ref{methods_ALE}.

Without first-order viscosity correction, the ALE method yields streaming patterns with comparable amplitude, vortex count, and rotation direction to those from RSM under the SD condition. In contrast, the NS-condition results exhibit opposite vortex directions and amplitudes differing by 1–2 orders of magnitude. Incorporating viscosity correction in the ALE method increases velocity amplitude without altering the vortex structure. Additionally, the ALE method reveals distinct velocity peaks near the bottom vortices adjacent to the sidewalls, a feature absent in SD-condition RSM results.

Numerical singularities arise near the lower corners of the channel due to discontinuities between the first-order velocity and impedance boundary conditions, consistent with the analysis by Barnkob~\textit{et al.}~\cite{barnkob_acoustically_2018}. These effects are more pronounced under the SD condition than the NS condition, but do not affect the streaming distribution away from the corners. A fully coupled three-dimensional piezoelectric model may help mitigate this issue.

\subsection{Semi-Analytical Slip Velocity Methods and Bulk Attenuation Effects}

To avoid the need for boundary-layer meshing, semi-analytical slip velocity methods have been developed for calculating boundary-driven streaming~\cite{lee_nearboundary_1989,vanneste_streaming_2011,barnkob_acoustically_2018}.

\begin{figure}[htbp!]
 \centering
 \includegraphics[width=0.45\linewidth]{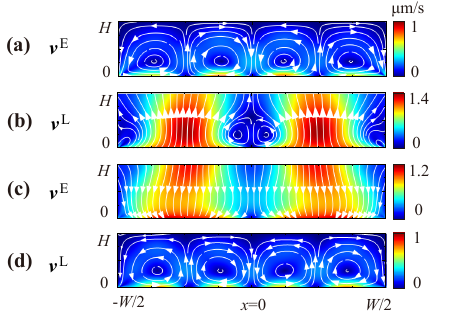}
 \caption{Boundary-driven streaming patterns from different semi-analytical slip-velocity methods: (a, b) Eulerian and Lagrangian velocities ($\bm{v}^{\mathrm{E}}$, $\bm{v}^{\mathrm{L}}$) from Lee \& Wang's method; (c, d) Eulerian and Lagrangian velocities from Bach \& Bruus's method. SSAW device parameters match Fig.~\ref{SSAW_discussion}.}
 \label{methods_LVM}
\end{figure}

In the limiting velocity method~(LVM), Nyborg~\cite{nyborg_acoustic_1958} and Lee \& Wang~\cite{lee_nearboundary_1989} adopted Eq.~(\ref{eqn-no_slip})~(the NS condition) as the second-order boundary condition. In contrast, the slip velocity derived by Vanneste~\textit{et al.}~\cite{vanneste_streaming_2011} and Bach~\textit{et al.}~\cite{bach_theory_2018} corresponds to Eq.~(\ref{eqn-stokes_drift})~(the SD condition). For flat vibrating surfaces, the expressions given by Vanneste and Bach are similar in form but differ in their coefficients. Detailed expressions are provided in Section K of the the Supplemental Material~\cite{supplementary}.

Figure~\ref{methods_LVM} shows boundary-driven streaming patterns for the SSAW device used in Fig.~\ref{SSAW_discussion}, computed using the methods of Lee \& Wang and Bach \& Bruus. With Lee \& Wang's slip velocity, the Eulerian velocity $\bm{v}^{\mathrm{E}}$ captures the vortex structure, whereas the Lagrangian velocity $\bm{v}^{\mathrm{L}}$ does not. The opposite holds for Bach's method---$\bm{v}^{\mathrm{L}}$ reflects the vortices, while $\bm{v}^{\mathrm{E}}$ does not. This distinction stems from the inclusion~(or exclusion) of the Stokes drift boundary condition.

Notably, a comparison of Figs.~\ref{SSAW_discussion}, \ref{methods_ALE}, and \ref{methods_LVM}(d) reveals that boundary-driven streaming alone cannot fully account for vortex amplitudes in the bulk. In this particular device, the magnitude of Eckart
streaming in the bulk is comparable to that of Rayleigh streaming near the bottom. Unlike in BAW devices, the acoustic intensity in SAW devices induces significant acoustic attenuation in the bulk, which also generates vortices and must be considered in modelling.

\section{CONCLUSION}
This study evaluates the influence of second-order boundary conditions, NS versus SD, on acoustic streaming patterns in SAW devices. Through combined numerical simulation and experimental validation, it is demonstrated that the SD condition yields results in close agreement with particle image velocimetry measurements, accurately predicting both the structure and magnitude of Lagrangian streaming in both SSAW and TSAW configurations. In contrast, the NS condition overestimates velocities by 1–2 orders of magnitude and often fails to reproduce key vortex features observed experimentally.

The Stokes drift velocity is shown to align with the acoustic intensity in the bulk region, while within the boundary layer, it exhibits a significant horizontal component due to viscous shear. As channel height increases, the influence of the Stokes drift diminishes, and the Lagrangian and Eulerian flow fields converge in the bulk. The SD condition produces results analogous to those obtained using a Navier slip model with large slip length, though it imposes a higher computational cost and reduced convergence rates in the RSM method.

It is concluded that the Stokes slip boundary condition is essential for physically accurate simulations of SAW-driven acoustic streaming. Its use may significantly improve predictions of particle trajectories and vortex formation, thereby providing a reliable foundation for the design and analysis of SAW acoustofluidic devices.

\bibliographystyle{unsrtnat}
\bibliography{BCs_SAW_streaming}  






\end{document}